# 3D-aCortex: An Ultra-Compact Energy-Efficient Neurocomputing Platform Based on Commercial 3D-NAND Flash Memories

Mohammad Bavandpour, Shubham Sahay, Mohammad Reza Mahmoodi, Dmitri B. Strukov

University of California, Santa Barbara, ECE Department

## ABSTRACT

The first contribution of this paper is the development of extremely dense, energy-efficient mixed-signal vector-by-matrix-multiplication (VMM) circuits based on the existing 3D-NAND flash memory blocks, without any need for their modification. Such compatibility is achieved using time-domain-encoded VMM design. Our detailed simulations have shown that, for example, the 5-bit VMM of 200-element vectors, using the commercially available 64-layer gate-all-around macaroni-type 3D-NAND memory blocks designed in the 55-nm technology node, may provide an unprecedented area efficiency of 0.14 $\mu m^2$/byte and energy efficiency of ~10 fJ/Op, including the input/output and other peripheral circuitry overheads. Our second major contribution is the development of 3D-aCortex, a multi-purpose neuromorphic inference processor that utilizes the proposed 3D-VMM blocks as its core processing units. We have performed rigorous performance simulations of such a processor on both circuit and system levels, taking into account non-idealities such as drain-induced barrier lowering, capacitive coupling, charge injection, parasitics, process variations, and noise. Our modeling of the 3D-aCortex performing several state-of-the-art neuromorphic-network benchmarks has shown that it may provide the record-breaking storage efficiency of 4.34 MB/mm², the peak energy efficiency of 70.43 TOps/J, and the computational throughput up to 10.66 TOps/s. The storage efficiency can be further improved seven-fold by aggressively sharing VMM peripheral circuits at the cost of slight decrease in energy efficiency and throughput.

## 1. INTRODUCTION

The Vector-by-Matrix Multiplication (VMM) is the most common operation in deep neural networks and many other computationally-intensive data and signal processing systems [1-6]. This fact is the motivation for the current intensive development of efficient VMM circuits and optimal architectures for their deployment in neuromorphic processors. So far, most VMM implementations are digital, with numerous commercial and experimental processor architectures developed in the last several years [7-14]. Their performance on VMM-heavy benchmarks is much higher than that of the standard CPUs, in part due to using low-precision operations, sufficient in particular for most neuromorphic tasks [15-17], including the most frequent inference function. However, digital approaches to the VMM task lead to relatively sparse design, which necessitates storing most of the synaptic weights off-chip, and as a result paying large performance penalty for memory access [18].

Due to the limited required precision, the digital implementations of the VMM may be challenged by mixed-signal (MS) circuits based on advanced analog-grade non-volatile memory devices, such as ReRAM [19-22], phase-change [23,24], and embedded floating-gate memories [25-32]. Indeed, prior work on such circuits has demonstrated the possibility of rather dramatic, orders-of-magnitude advantages in energy, speed, throughput, and circuit density, over their digital counterparts [18,19,26,28].

However, the mixed-signal approaches to the VMM tasks have their own challenges. The developed technologies for fabrication of highly scalable emerging memristive devices are not yet mature, still requiring a substantial (orders-of-magnitude) improvements in device-to-device uniformity, and in device current reduction. The floating-gate memory cells, whose optimal design [33] mitigates these problems, have relatively large cells, even if implemented by re-design of highly optimized commercial flash memories [25]. The resulting relatively low circuit density may lead, just like in the case of the digital implementations, to significant inter- and intra-chip data transfer overheads [18]. Additional concern is large area/energy overhead of conversion between analog and digital domains in MS inference accelerator architectures.

These challenges have provided the main motivation for our work - the development of VMM circuits and architectures based on 3D-NAND memories [34-38]. Indeed, even the already developed commercial 3D-NAND memory technology enables record-breaking effective bit density, ultra-low fabrication cost per bit, and multi-level cell programming capability [37], while still advancing rapidly.

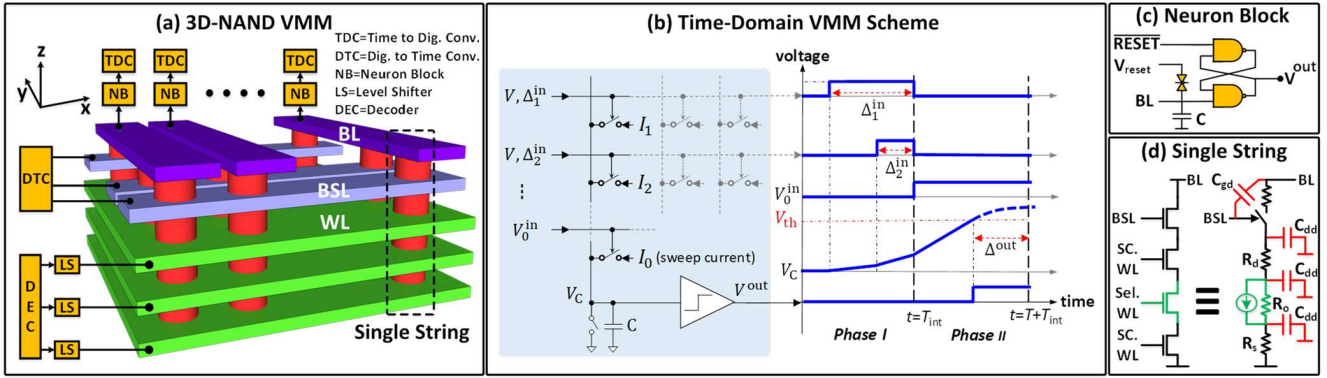

**Fig.1: The main idea of the 3D-VMM circuit. (a) Cartoon of 3D-NAND flash memory block and its use in the proposed circuit. For simplicity, a layer of transistors at the bottom of the block, which connects the cell strings to the common source (ground) is not shown. (b) Basic structure and example of operation in the utilized time-domain approach [27]. (c) Circuit diagram of the peripheral neuron, which consists of a load capacitor $C$, connected to the bit line (BL), and an SR latch, implementing a unit step function of its input. (d) Equivalent circuit of a single string for the operation mode.**

The specific contributions of this work are the following:

- A novel time-domain implementation of MS-VMM using commercial 3D-NAND flash memory blocks, without the need for its modification.
- A detailed analysis of non-idealities impacting compute precision in 3D-NAND VMMs and optimization of peripheral circuits to achieve target precision.
- A detailed discussion of baseline 2D-aCortex architecture focused on its unique features
- 3D-aCortex architecture and its enabling features (e.g. folded buffer, output accumulator, addressing scheme, multi-unit controller for efficient time-multiplexing).
- A 3D weight-packing algorithm, and memory requirement analysis process for 3D-aCortex
- A detailed performance results and their breakdowns for common deep-learning benchmarks.

Fig. 1a shows a typical 3D-NAND memory architecture. In it, many layers of memory cells are stacked on top of each other, with the cells connected in the $z$-direction (normal to the chip surface) to form a "string". On the top of each string, there is a bit-select-line (BSL) transistor that connects it to the bit line (BL). The memory block consists of a 2D ($x$-$y$-plane) mesh of such strings, with all memory cells of the same level (i.e., at the same $z$-position) sharing the common word-line (WL) metal plate. In addition, the strings share BSLs in the $x$-direction, and BLs in the $y$-direction.

While showing a possible dramatic increase of the stored weight density (scaling as the number of the cell layers), this figure also indicates a major problem for the VMM implementation. Namely, sharing of each word line by all cells of that layer does not allow to use the "current-mode" approach [28, 29] that was successfully employed [18, 25] for the adaptation of a commercial 2D flash memory for MS-VMM. In future, an appropriate redesign of the 3D wiring (perhaps, as in the 2D work [18, 25], not touching the highly optimized memory cells) may be the best option. However, such modification (assumed in the recent work [39]) would require a major technological effort. (Additionally, the approach [39] requires using high-resistance and high-capacitance WL on the critical path).

In this work, we have shown that the time-domain approach to the VMM function [27, 40-42] may enable using commercial 3D-NAND memories without any modification. After describing this approach in the beginning of section 2, we use the balance of that section to describe the methods of our detailed, quantitative analysis of the possible performance of the resulting 3D-VMM blocks, taking into account various non-idealities impacting their performance.

## 2. 3D-VMM DESIGN

### 2.1. Time-Domain VMM

The target analog VMM operation may be represented as

$$y_j = \frac{1}{M}\sum_{i=1}^{M} w_{ij}x_i,\qquad(1)$$

where $x_i$, $w_{ij}$, and $y_j$ are real numbers, which may take any values on the [0, 1] segment. In the time-domain approach [27], the components $x_i$ and $y_i$ of the input and output vectors are encoded with the durations $\Delta$ of fixed-amplitude pulses: $\Delta_i^{in} = x_iT$, $\Delta_j^{out} = y_jT$, where $T$ is a certain fixed time window, while the matrix elements ("weights") $w_{ij}$ are represented by adjustable current sources $I_{ij}$ within a fixed range [0, $I_{max}$]: $w_{ij} = I_{ij}/I_{max}$. (In floating-gate memory cells, the weights are kept in the form of stored floating gate charges, which define the source-to-drain currents $I_{ij}$ at a fixed drain voltage.)

The computation is performed in two phases (Fig. 1b). During the first $T_{int}$-long (integration) phase, the input pulse $\Delta_i^{in}$ turns on fixed drain voltages, and hence the current sources $I_{ij}$ of the $i$th row, leading to the injection of electric charges equal to $I_{ij}\Delta_i^{in} \propto w_{ij}x_i$ into the $j$th column through the corresponding memory cells. The charges from multiple rows of the $j$th column are summed up on its load capacitor $C$. As a result, by the end of phase I, the capacitor voltages $V_C$ (which are reset before the operation) become proportional to the component of the desired VMM output vector:

$$V_{C,j} = \frac{1}{C} \sum_{i=1}^{M} I_{ij} \Delta_i^{in}. \qquad (2)$$

During the second $T$-long phase, these voltages are converted into the durations $\Delta_j^{out}$ of the output pulses (Fig. 1b). This is done by additional charging of each load capacitor with a constant "sweep" current equal to $M I_{max}$, inducing a linear ramp-up of its voltage in time, starting from the value (2). At the moment when the total voltage reaches the fixed threshold $V_{th}$, an output fixed-amplitude pulse is initiated, with its falling edge aligned with the end of this phase II. As a result, the duration of the output pulse generated in phase II is

$$\Delta_j^{out} = \frac{1}{M I_{max}} \sum_{i=1}^{M} I_{ij} \Delta_i^{in}. \qquad (3)$$

where, just for convenience, all load capacitances are assumed to be equal to $C = M I_{max} / V_{th}$. Also, note that $T \geq T_{int}$.

The described approach can be easily extended to four-quadrant time-domain VMM, by using differential rows/columns, and a set of four cells for each weights, to represent positive and negative inputs/outputs [34].

### 2.2. 3D-VMM Structure and Operation

In our 3D-VMM block, each elementary ("single-shot") VMM operation uses the weights recorded in the floating-gate cells of one $x$-$y$ layer of the 3D-NAND memory circuit (see Fig. 1a again). This layer is selected by setting its word line (WL) voltage to 2 V, while setting the cells of all other layers to the highly conductive "pass" state by applying 5 V to those word lines. The cell currents are collected and integrated at the bit lines (BL). However, irrespective of the selected layer of cells, the inputs are always applied to bit-select lines. The "sweep" currents, necessary for phase II of the operation, are injected through the top layer of cells of all strings, enabled by a positive voltage applied to all bit-select lines (BSL).

Such elementary VMM operations, based on different layers, are used as steps of the time-division-multiplexing operation. Clearly, such VMM operation mode does not require changes in the usual NAND flash memory array, and only needs to complement it with custom-designed peripheral decoder and level-shifter circuits.

Note that because of significant WL parasitics in 3D-NAND memory, the total delay for performing one VMM elementary operation is $2T_{LS} + T_{int} + T$, where $T_{LS}$ is the time required to select a certain layer.

### 2.3. Non-Idealities

For our detailed analysis, we have specifically considered the 3D-NAND memory based on polysilicon gate-all-around macaroni-body charge-trap cells. Besides its widespread use, another reason for this choice is availability of a behavioral compact model for such memory, which may be used for quantitative simulation. In such model, individual cells are approximated as cylindrical gate-all-around nanowire FETs with a voltage-controlled-current-source [39]. The model takes into account various parasitic capacitance coupling effects, and accurately reproduces the experimental string current characteristics [43, 44].

We next discuss the most important factors affecting computing precision:

#### A. Drain-Induced Barrier Lowering

Let us first note that since the current is sunk through the cells to the source line, we consider the scheme in which BL voltage is charged to a voltage $\Delta V_D + V_{th}$ at the start of phase I, where $\Delta V_D$ is the total voltage swing on BL during computation, and then discharged to $V_{th}$ in the phase II.

DIBL error is defined as a relative difference of currents via string of cells at two extreme BL voltages, i.e.

$$E_{DIBL} \simeq 1 - I(V_{th})/I(V_{th} + \Delta V_D). \qquad (4)$$

Without considering additional headroom to deal with capacitive coupling, the typical values are $V_{th} = 0.6$ V and $\Delta V_D = 0.2$ V, which correspond to the quasi optimal conditions for the CMOS-based neuron implementation [12].

According to Eq. 4, the DIBL error is proportional to the small signal transconductance gain $\delta I_D / \delta V_D$ of a string over the target operating regime. Given the small signal model shown in Fig. 1d, the transconductance gain can be formulated as:

$$\frac{\partial I_D}{\partial V_D} = \frac{1}{R_D + R_0 + (1 + g_m R_0) R_S}, \qquad (5)$$

where $g_m$ and $R_0$ are the small signal parameters of a single memory cell, and $R_D$ and $R_S$ are the lumped string resistances on the drain and source side, respectively, of the selected memory cell. According to Eq. 5, larger $R_D$ and $R_S$ help reducing the DIBL error, but at the cost of limiting the current range. Moreover, because of stronger effect of $R_S$, DIBL error is less for top memory cells (which was the reason for using top layer for sweep currents). Also, DIBL error is less for larger string currents due to intrinsically larger $R_0$, when the selected cell operates closer to strong inversion mode. These observations are confirmed by modeling (Fig. 2). In line with Eq. 4, DIBL error increases almost linearly with the total swing in the target operation region (Fig. 2b).

#### B. Capacitive Coupling

Due to the switched-capacitor nature of the proposed approach, capacitive coupling is a significant source of compute error. We break down the sources of coupling into two components. The first component, gate-drain (GD) coupling, is caused by their overlap in BSL transistor and coupling between BSL and BL wires. The second one (DD) is caused by the parasitic capacitors between the string and the rest of the memory block. These two lumped capacitors are denoted as $C_{gd}$ and $C_{dd}$, respectively (Fig. 1d).

Note that $C_{dd}$ is distributed over the total length of the string. When a 2.5 V rising edge is applied to BSL line, GD coupling results in an immediate positive disturbance charge on the BL voltage with the amount of $C_{gd} \times (2.5 \text{ V})$. Moreover, when the string is selected via BSL, DD coupling causes a negative disturbance charge on BL to charge the string parasitic capacitors $C_{dd}$ from their initial voltage (ground) to their final DC voltage at which the string sinks the target current. When a 2.5 V falling edge is applied to BSL, the capacitive coupling is dominated by the GD coupling which causes an immediate negative disturbance charge on BL by $-C_{gd}\times(2.5 \text{ V})$.

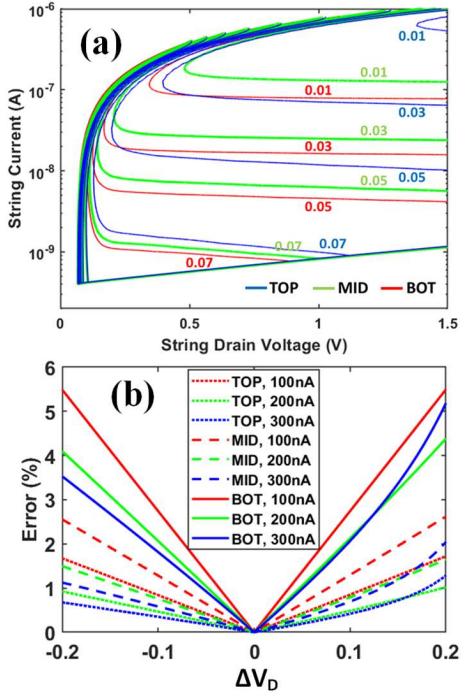

**Fig. 2: (a)** Small-signal DIBL error contours (shown in %) in $I_D$-$V_D$ space for top, middle, and bottom layer memory cells, programmed in various states in a 64-layer 3D-NAND memory. Small-signal error is defined as $100\times(1 - I(V_D) / I(V_D+1 \text{ mV}))$, i.e. relative change in string current for a 1 mV change in the BL voltage. **(b)** Total DIBL error (%) for $\pm0.2$ V swing on the drain voltage around $V_D = 0.7$ V for various memory states.

GD coupling disturbance is almost independent of the selected cell location and programming state, while the DD coupling disturbance during rising edge is highly dependent on both (Fig. 3). The amplitude and time constant of the DD charge disturbance are both larger for the cells closer to the bottom of the string due to higher voltage variation on the parasitic capacitors ($C_{dd}$), especially the ones closer to the bottom but higher than the selected cell where the path to both ground and BL are highly resistive.

Taking into account the coupling, we can formulate the amount of voltage disturbance on the BL for each input as $\Delta V_{cp} = Q_D/C_0$ where $C_0$ is the amount of load capacitance per input, and $Q_D$ is the total disturbance charge caused by one input in both phase I when the target weight layer is selected and a rising edge followed by a falling edge is applied to BSL, and also phase II when the sweeping layer, i.e. top layer, is selected and one rising edge is applied to BSL. A major portion of $Q_D$, and consequently $\Delta V_{cp}$ is dependent on the location of target weight layer (Fig. 3b). Hence the maximum disturbance charge $(Q_D)_{max}$, which causes the largest disturbance voltage swing on BL $(\Delta V_{cp})_{max} = (Q_D)_{max}/C_0$, occurs when the target weight layer is at the bottom of the string.

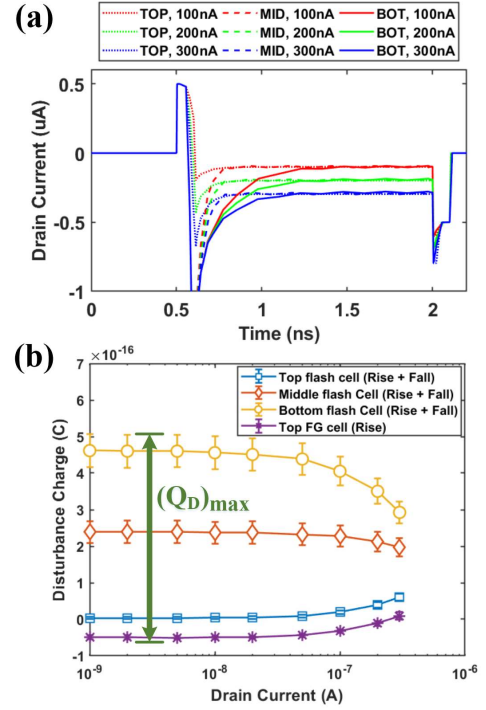

**Fig. 3: Charge disturbance on BL due to capacitive coupling.** **(a)** Time domain representation of drain (BL) current and its disturbances caused by coupling when a 2.5V rising edge (at $t = 0.5$ ns) followed by a same-amplitude falling edge (at $t = 2$ ns) is applied to the BSL for various programming states where the selected cell is located at top, middle, and bottom layer of the string. **(b)** Total string disturbance charge on a drain caused by capacitive coupling when a 2.5 V rising + falling edge is applied to BSL and target cell is located in top, middle, and bottom layer and programmed in various states (corresponding to phase I of computation), as well as when a single 2.5 V rising edge applied to BSL and target cell is located in top layer and programmed in various states (corresponding to phase II of computation). Error bar represents 3$\sigma$ distribution of the disturbance charge due to process variations.

In order to support VMM operation on all the layers, reset voltage $\Delta V_D + V_{th}$ should be selected to reserve a portion of total voltage swing on BL for the worst case voltage variation due to coupling. Hence, we select $\Delta V_D = \Delta V_{cmp} + (\Delta V_{cp})_{max}$, where $\Delta V_{cmp}$ is the voltage swing without considering the capacitance coupling for the weight and sweep current sources. Though the utilized differential scheme is robust to coupling, the output time window in which the output pulse is generated should be scaled by a coupling coefficient $\alpha_{cp} = 1 + (\Delta V_{cp})_{max}/ \Delta V_{cmp}$. Note that a small portion of $(\Delta V_{cp})_{max}$ still affects the output precision because of difference in disturbance charge caused by positive and negative sub-weights due to process variation, and dependence of disturbance charge on the programmed state of the flash cells. Note that a larger $(\Delta V_{cp})_{max}$ also leads to a higher BL voltage swing and consequently a larger DIBL error.

| Input time window $T_{int}$ | 8 ns | | | 16 ns | | | 32 ns | | |
|---|---|---|---|---|---|---|---|---|---|
| Maximum cell current $I_{max}$ | 100nA | 200nA | 300nA | 100nA | 200nA | 300nA | 100nA | 200nA | 300nA |
| Load capacitor per input $C_0$ (fF) | 4 | 8 | 12 | 8 | 16 | 24 | 16 | 32 | 48 |
| Coupling vol. swing $\Delta V_{cp}^{max}$ (mV) | 150 | 75 | 50 | 75 | 32.5 | 25 | 32.5 | 16.25 | 12.5 |
| Coupling coefficient, $\alpha_{cp}$ | 1.75 | 1.375 | 1.25 | 1.375 | 1.1875 | 1.125 | 1.1875 | 1.094 | 1.062 |
| Output time window $T_{out}$ (ns) | 14 | 11 | 10 | 22 | 19 | 18 | 38 | 35 | 34 |
| Single device $SNR^{cell}$ (dB) | 33.97 | 36.98 | 38.75 | 36.98 | 40 | 41.76 | 40 | 43.01 | 44.77 |
| Single device noise 3σ error (%) | 12 | 8.48 | 6.92 | 8.48 | 6 | 4.89 | 6 | 4.24 | 3.46 |
| Noise-free VMM comp. error (%) | 6.24 | 3.55 | 1.79 | 4.25 | 2.31 | 1.16 | 3.62 | 1.92 | 0.96 |
| Final compute error $M$ =10 (%) | 10.03 | 6.23 | 3.98 | 6.93 | 4.20 | 2.71 | 5.51 | 3.26 | 2.05 |
| Final compute error $M$ =100 (%) | 7.44 | 4.40 | 2.48 | 5.10 | 2.91 | 1.65 | 4.22 | 2.34 | 1.30 |
| Final compute error $M$ =1000 (%) | 6.62 | 3.81 | 2.01 | 4.52 | 2.50 | 1.31 | 3.81 | 2.05 | 1.07 |

**Table 1: Design space exploration. circuit specs and compute error (due to noise and circuit nonidealities) for various choices of $T_{int}$ and $I_{max}$. final VMM error is reported for three different VMM sizes ($M$ = 10, 100, and 1000), and the achievable output bit-precision is shown by a color coding scheme in which orange = 2 bits, blue = 3 bits, green = 4 bits, and yellow = 5 bits.**

*C. Noise*

White (shot/thermal) noise will dominate at the considered high-bandwidth operation. (We assume that the cells with extremely high flicker noise will be set to high conductive states and avoided during mapping.)

The noise power for a single string operating in subthreshold can be approximated as $\sim 2qI_{max}/T$, while SNR for a single device as $SNR^{cell} \approx 2q/I_{max}$, where $q$ is an electron charge. Accordingly, for an $M \times 1$ VMM unit (a dot product), noise and signal power are $P_{noise}^{M \times 1} = \frac{2qMI_{max}}{T}$ and $P_{signal}^{M \times 1} = (MI_{max})^2$, respectively. Hence,

$$SNR^{M \times 1} = \frac{P_{signal}^{M \times 1}}{P_{noise}^{M \times 1}} \approx \frac{MI_{max}T}{2q} = M \times SNR^{cell}. \qquad (6)$$

The equivalent 3σ error due to noise is derived as

$$E_{3\sigma}^{M \times 1} \approx \frac{2 \times 3 \times \sqrt{\frac{2qMI_{max}}{T}}}{MI_{max}} = 6 \times \sqrt{\frac{2q}{MI_{max}T}} = \frac{E_{3\sigma}^{cell}}{\sqrt{M}}. \qquad (7)$$

Note that in the above equation, the distribution is multiplied by two due to the differential scheme. According to the derived equation, compute error is inversely proportional to the square root of maximum current, compute time window, and the VMM size.

## 2.4. Computing Precision

The compute (output) precision $p_O$ can be defined separately from the weight precision [27] as

$$p_O = -\log_2(E_C) - 1, \quad E_C = \frac{1}{T}\max_{\Delta^{out}}|\Delta^{ideal} - \Delta^{out}|, \qquad (8)$$

where $E_C$ is a maximum absolute difference between the ideal ($\Delta^{ideal}$) and actual ($\Delta^{out}$) output pulse durations, normalized by its maximum value.

The 3D-VMM circuit can be designed following various optimization targets such as precision, energy, speed and area. Here, we focus on the precision which generally limits the design space in application-specific hardware design. The main tunable circuit parameters impacting our 3D-VMM's precision are $I_{max}$ and $T_{int}$.

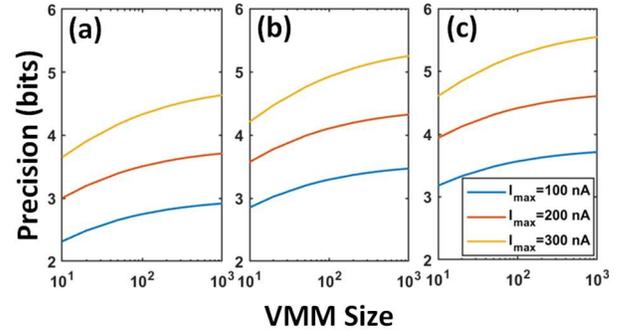

**Fig. 4: 3D-NAND based VMM bit precision with respect to VMM size for $I_{max}$ = 100 nA, 200 nA, and 300 nA for $T_{int}$ = (a) 8 ns, (b) 16 ns, and (c) 32 ns.**

In Table 1, various combinations of $(T_{int}, I_{max})$ are targeted to investigate the impact of these parameters on 3D-VMM's compute precision. Assuming $\Delta V_{cmp} = 0.2$ V and $(Q_D)_{max} = 6 \times 10^{-16}$, we first calculate dependent parameters such as load capacitor, coupling voltage disturbance, and output time window for every combination of $I_{max}$ and $T_{inp}$. Then, full circuit-level SPICE simulations are performed on 10 different VMM sizes from $10 \times 10$ to $1000 \times 1000$ with 1000-times randomized inputs/weights considering detailed parasitic models for the interconnect wires and devices, and also process variations considering the 55-nm technology node. The results for different simulated scenarios show that the compute error for the noise-free circuit remains relatively constant over the target VMM size range.

Table 1 also reports the SNR and 3σ noise error parameters, calculated according to Eqs. 6 and 7, and total error targeting three representative VMM sizes. Fig. 4 shows that bit-precision, corresponding to the calculated error, increases with respect to $I_{max}$, $T_{int}$, and VMM size.

## 2.5 Weight Precision

Similar to 2D flash memory circuits [25, 26], the weight precision in 3D-VMM is also expected to be affected by the tuning accuracy and drift of the analog memory state. The additional challenge for cell current tuning will be relatively large resistance $R_D$ and $R_S$ (Fig. 1d). The voltage drops across these resistors (especially $R_S$) must be taken into account while optimizing the programming scheme for a target output current.

Quantitative analysis of such factors is challenging, mostly due to the lack of published relevant data. It should be noted, however, that the utilization of barrier-engineered materials and the gate all-around architecture in the 3D-NAND memory results in a narrower threshold voltage distribution and a lower threshold voltage shift due to cell-cell coupling as compared to the planar counterparts. In fact, multi-level state capabilities (> 3-bits) have been routinely demonstrated in 3D-NAND memories, and is expected to further improve as its technology continues to advance [35-37].

## 3. 3D-NAND VMM RESULTS

As was described in the last section, the 3D-VMM parameters can be chosen to operate with any precision from 2 bits to 5 bits. In this section, we describe the results obtained for the 4-bit precision, which has been proved to be sufficient for most tasks of neuromorphic computations [15-17]. A 4-bit 3D-VMM block consists of the following main components (Fig. 1a):

- **DTC** converts the digital input to the time-domain pulse of fixed amplitude and controllable duration. As was described earlier [27], this unit includes one shared 4-bit counter and one 4-bit comparator connected to a 1-bit latch per input.

- **3D-FM** is the 3D-NAND memory block for the $M \times N$ (per layer) VMM, which consists of $M \times 2N$ cells with the dimensions reported in [38,43], as well as an extra marginal space for routing the word and bit-select lines. Note that the parasitics of the word-line plate extensions by routing and vias/wires are taken into due account in the simulations.

- **CAP** stands for the load capacitor. Here we assume that it is implemented as MOSCAP in the 55-nm technology, and also account for an extra marginal space around each capacitor. (Using MOM and MIM capacitors should further improve the results.)

- **NB** represents the neuron circuit, consisting of a pair of NAND latches and a couple of AND and NOT logic gates for implementing the differential scheme.

- **TDC** converts the time-encoded digital output to the corresponding digital output number. This unit consists of a 4-bit adder and a 4-bit DFF per output. The adder and the DFFs are connected to form an accumulator, counting the duration of the output pulse, using clock pulses which are shared by all accumulators. Note that this unit along with DTC constitutes the "I/O".

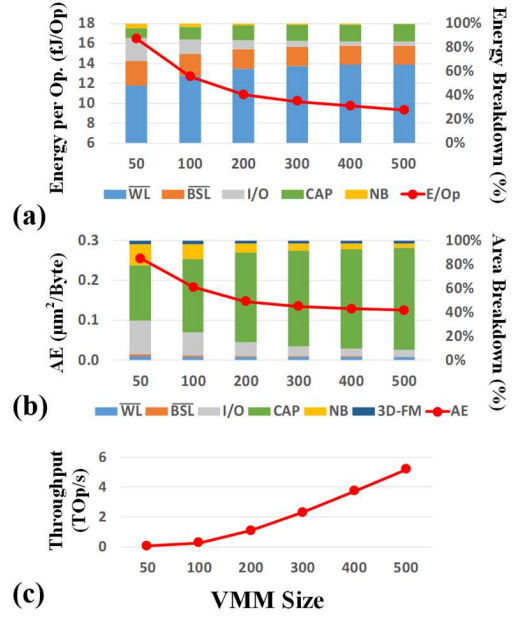

**Fig. 5: 3D-NAND based VMM performance metrics. (a) Energy per operation breakdown. (b) Area efficiency breakdown. (c) Throughput as a function of VMM size.**

- $\overline{WL}$ represents the word-line level shifters, which apply the read/pass voltages (2 V / 5 V) to the word-line plates (Fig. 1a). Note that the width of each driver transistors is made proportional to the area ($M \times N$) of the plate it serves, in order to keep the layer selection time ($T_{LS}$) within a limited range comparable to the computation time.

- $\overline{BSL}$ is an array of level-shifters driving the bit-select lines and converting the 1.2 V time-encoded, fixed-amplitude inputs back to 2.5 V digital pulses.

As Table 1 shows, the optimal design point, which guarantees the 4-bit precision across VMMs of various size is $\{I_{max} = 300 \text{ nA}, T_{int} = 16 \text{ ns}\}$. Fig. 5 shows the energy, area, and throughput calculation results for various sizes of our 3D-VMM, as well as the energy and area breakdowns for this design point. According these results, the energy consumption is dominated by the word line selection and by feeding the inputs into the bit-select lines, despite the fact that their capacitance (per cell) is lower than the load capacitance $C_0$ (The reason is a higher voltage swing on these lines). Moreover, the contribution of I/O and neuron circuits into the energy consumption (per operation) decreases as the VMM size increases, due to their higher sharing factor. As a result, the energy per operation is only ~9 fF for $M = N = 500$.

Fig. 5b shows the area breakdown and the area per synapse (i.e. per weight). The area is highly dominated by the CAP, though its contribution is minor in energy consumption. Moreover, the share of I/O and neuron in the area per operation also decreases when the size of VMM is increased, due to enhancement in their sharing factor. Finally, Fig. 5c shows the VMM's throughput for its various sizes considering scaling to maintain $T_{LS}$ within the range of [20 ns, 30 ns].

The results show that the proposed 3D-VMM achieves a ~100× better area efficiency than that of its 2D-NOR memory-based counterpart [27], while maintaining a comparable energy efficiency and throughput. Such high area efficiency of our 3D-VMM enables its efficient system-level deployment via minimizing the data transfer overhead - see the next section.

# 4.  aCORTEX

## 4.1  2D-aCortex Architecture

aCortex is MS NVM-based neuromorphic inference accelerator, which is specifically designed to minimize peripheral circuitry overhead maximize by performing more computation and communication in the analog domain. Fig. 6a shows the overall structure of 2D-aCortex architecture. Its main components, for processing a $K$-word data stream (with $p$ bits per word), are:

(1) a central eDRAM-based main memory (MM) with one input and one output port of $K$-word width,

(2) a set of configurable local $K$-word buffers supporting both individual load and load & shift operations with flexible chain size,

(3) a 2D mesh of processing elements (PEs) each including a core $K \times K$ analog VMM,

(4) a set of integrate-digitalize unit (IDU) blocks including $K$ neurons, ADCs, and activation functions,

(5) an auxiliary block (AUX) including an array of $K$ digital comparators/adders/multipliers to perform infrequent neuromorphic operations such as max-pooling and element-wise vector addition and vector-by-vector multiplication in digital domain, and finally (6) a controller including an instruction memory (IM) and digital circuitry to produce control signals for other blocks.

Input data are loaded from the MM into buffers through a shared $K$-word digital load bus (L-Bus). This stream is vertically propagated from buffers into the PEs through a shared $K$-word input bus (I-Bus) in analog (or digital) domain where the input data conversion is done globally (or locally) in the buffer (or PE) blocks. Analog outputs of the PEs are integrated on the shared analog output bus (O-Bus) and converted back to digital domain in the IDU blocks, in which the activation function is also applied. The final output data is stored back into the MM via a shared $K$-word digital store bus (S-Bus).

The VMM operator of variable size can be implemented by enabling target PEs on which the weights are pre-programmed, as well as their corresponding buffers and IDUs, while the rest of the blocks are disabled/inactive (Fig. 6a). Note that an output re-scaling scheme in the PE/IDU might be needed (especially when PE output is current-mode) in order to handle various input sizes. Accordingly, multiple

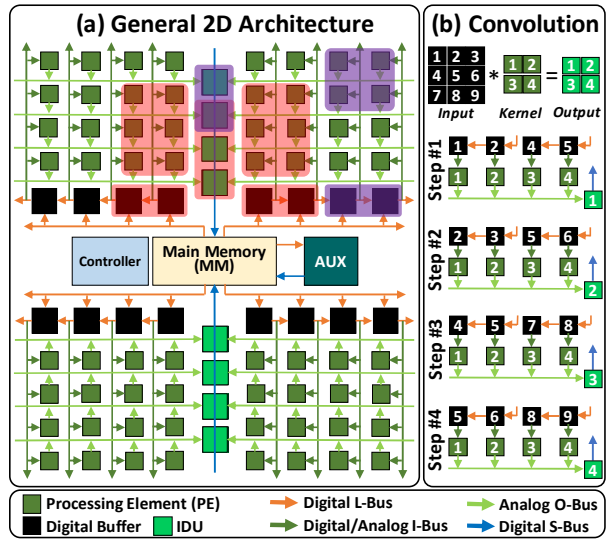

**Fig. 6:** (a) 2D-aCortex architecture for a general-purpose neuromorphic processor employing a nonvolatile memory-based analog VMM as the core processing element (PE). (b) Convolution operation on the 2D-aCortex processor.

VMM kernels of various dimensions may be packed into the 2D structure of PEs and utilized one at a time.

The 2D-aCortex performs inference tasks in a layer-by-layer manner, by storing intermediate data in the MM. A full connection between two layers may be performed in a single VMM cycle. In contrast, the convolutional connection is performed one output pixel at a time in the row-first manner as shown in Fig. 6b. For that, one copy of the convolutional kernel is pre-programmed into a specific location in the 2D structure of PEs, and then activated in multiple steps to calculate different output pixels. Such scheme, along with the reconfigurable buffers, allows efficient data reuse. The recurrent-layer function is also performed in multiple steps, generally by loading the output of the previous step, along with its corresponding element of the input sequence, into the buffers, and activating the same kernel until the sequence has been finished. Note that according to this scheme, any network with various interconnections of the discussed basic layers, such as Residual layers [5], Inception layers [4], Bi-directional and Residual Recurrent layers [6], etc. may be performed on this architecture - as long as the MM is large enough to keep the intermediate data.

## 4.2  3D-aCortex Architecture

The architecture of the proposed 3D-aCortex is derived from that of the 2D-aCortex, using the general transformation scheme shown in Fig. 7a. Indeed, the 2D-aCortex is equivalent to a very large VMM operator in which the digital inputs are read into the buffer blocks (shown black), which can be configured as shift registers to minimize the need in the MM access at convolution tasks. The inputs are converted into analog/time-domain signals and propagated to vertical input lines of the 2D NVM array, while analog output signals, aggregated on the shared output lines of the array are converted back to digital values and stored in local buffers (shown green), waiting to be written into the MM.

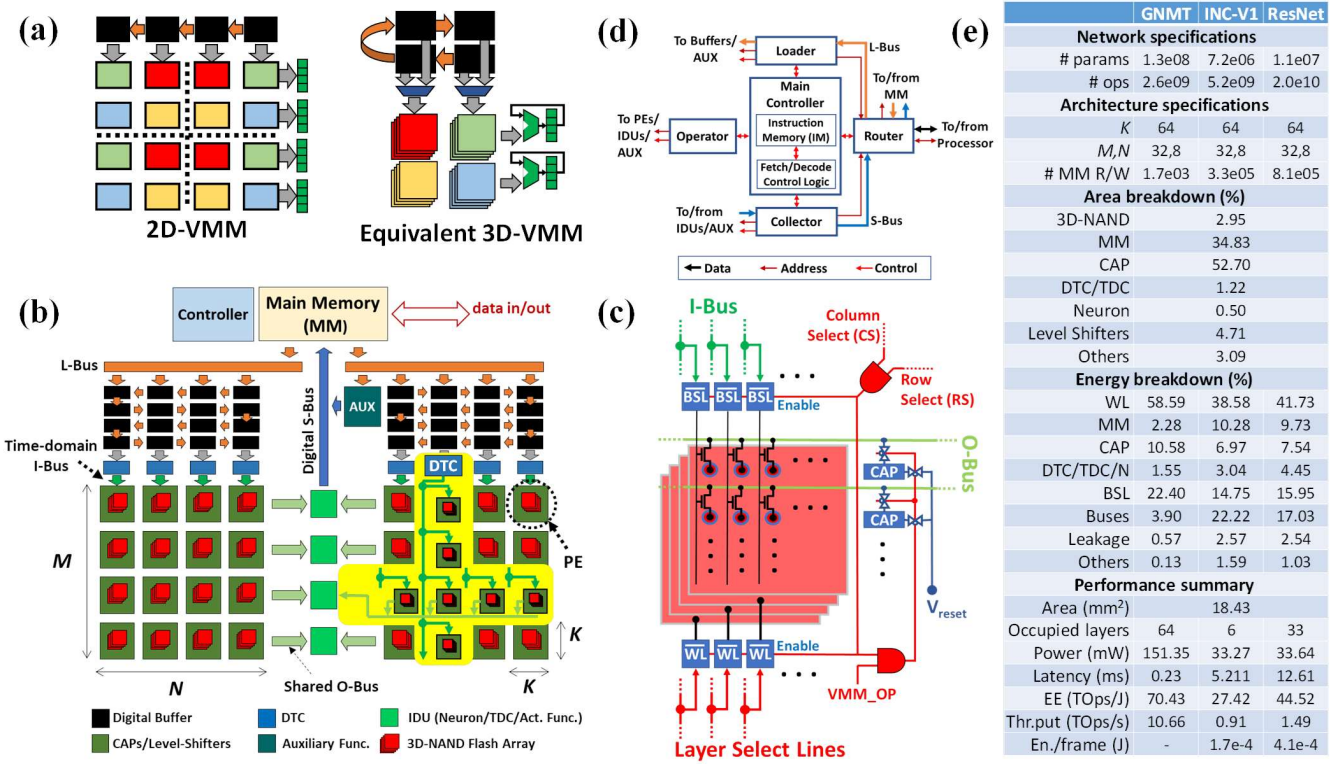

| | GNMT | INC-V1 | ResNet |
|---|---|---|---|
| **Network specifications** | | | |
| # params | 1.3e08 | 7.2e06 | 1.1e07 |
| # ops | 2.6e09 | 5.2e09 | 2.0e10 |
| **Architecture specifications** | | | |
| K | 64 | 64 | 64 |
| M,N | 32,8 | 32,8 | 32,8 |
| # MM R/W | 1.7e03 | 3.3e05 | 8.1e05 |
| **Area breakdown (%)** | | | |
| 3D-NAND | 2.95 | | |
| MM | 34.83 | | |
| CAP | 52.70 | | |
| DTC/TDC | 1.22 | | |
| Neuron | 0.50 | | |
| Level Shifters | 4.71 | | |
| Others | 3.09 | | |
| **Energy breakdown (%)** | | | |
| WL | 58.59 | 38.58 | 41.73 |
| MM | 2.28 | 10.28 | 9.73 |
| CAP | 10.58 | 6.97 | 7.54 |
| DTC/TDC/N | 1.55 | 3.04 | 4.45 |
| BSL | 22.40 | 14.75 | 15.95 |
| Buses | 3.90 | 22.22 | 17.03 |
| Leakage | 0.57 | 2.57 | 2.54 |
| Others | 0.13 | 1.59 | 1.03 |
| **Performance summary** | | | |
| Area (mm²) | 18.43 | | |
| Occupied layers | 64 | 6 | 33 |
| Power (mW) | 151.35 | 33.27 | 33.64 |
| Latency (ms) | 0.23 | 5.211 | 12.61 |
| EE (TOps/J) | 70.43 | 27.42 | 44.52 |
| Thr.put (TOps/s) | 10.66 | 0.91 | 1.49 |
| En./frame (J) | - | 1.7e-4 | 4.1e-4 |

**Fig. 7:** Baseline 3D-aCortex architecture and results, with no CAPs sharing among 3D NAND flash blocks. (a) 2D to 3D architecture transformation. (b) 3D-aCortex architecture layout and main blocks. (c) Controller sub-blocks and signaling between them. (d) PE's main circuit components and control circuitry. (e) Performance estimates for GNMT [6], Inception-v1 [4], and ResNet [5].

The 3D equivalent of such a 2D VMM operator is shown on the right panel of Fig. 7a, which assumes a multi-step (here 4-step) VMM operation, at which each weight sub-matrix is selected in one step. In order to avoid an increase in the number of MM accesses per VMM operation during the 2D-to-3D transformation, the input/output blocks should be redesigned. At the input, the shift registers are folded, and an extra selector is added to discriminate between vertically aligned buffer blocks at various VMM steps. Moreover, at the output, a digital accumulator with extra precision is added, in order to temporally aggregate the partial results. Finally, the most significant part of the result is selected for the final output, using a barrel shifter to match the target precision.

Following such transformation scheme, we have proposed the 3D-NAND based DNN/RNN processor architecture shown in Fig. 7b. Its main components are:

**PE:** In this architecture, PEs are placed as a $M \times 2N$ 2D structure where they share time-domain inputs in vertical direction (I-Bus), and analog BL output in horizontal direction (O-Bus). As shown in Fig. 7c, each PE includes a core 64-layer 3D-NAND memory with the size of $K \times 2K$ and also the peripheral circuitry. The peripheral circuitry for each PE includes: 1) $K$ local load capacitors (**CAP**) connected to the shared BLs and also $V_{reset}$ through pass gates, 2) 64 WL level-shifters ($\overline{\text{WL}}$) and drivers for selecting the target layer, 3) $K$ BSL level-shifters ($\overline{\text{BSL}}$) and drivers for changing the voltage level of the shared time-domain input, and also activating the inputs during the phase II of computation, and 4) control logic gates for enabling/disabling the unit components.

The column select (CS) and row select (RS) lines are propagated respectively in vertical and horizontal directions to select and enable the target PEs. Moreover, the **CAP** pass-gates in the enabled PEs are set to VMM operation mode at the appropriate time through a control signal called VMM_OP.

**IDU:** Each IDU block includes three sub-blocks as: 1) neuron latches receiving input from O-Bus, 2) TDCs which are digital accumulators with higher precision (here 6-bit where 2 extra bits enables accumulating results for VMM operation on 4-layers, i.e. $4 \times 2N \times K$ inputs without overflow), 3) barrel shifters to select the target output bit locations, and 4) activation function circuitry which applies a target nonlinear function (here linear, ReLU, tanh, or sigmoid) to the TDC's output.

**Controller:** Due to the flexibility of 3D-aCortex, any VMM operation up to $MK \times NK$ can be performed in one VMM step. In order to perform a desirable size VMM on one layer of the 3D-NAND memory: 1) target PEs and their corresponding DTC and IDU units are enabled, 2) input data is loaded into the buffers while target 3D-NAND memory layer is selected, 3) enabled PEs are set to VMM operation mode and DTCs convert and apply time-domain inputs during Phase I of operation, 4) sweeping (first) layer in 3D-NAND memory is selected, 5) outputs are calculated, converted back to digital, and finally 6) the results pass through the target activation function and stored in the MM.

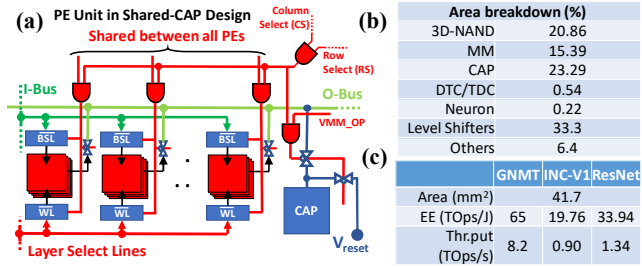

**Fig. 8: 3D-aCortex with CAP sharing: (a) main changes in PE design with respect to the baseline architecture, (b) area breakdown, and (c) performance estimates.**

Since VMM area is largely dominated by load capacitors (Fig. 5b), we have also investigated a variation of the baseline 3D-aCortex architecture in which a CAP circuit is shared among 16 3D-NAND blocks, each with its own BSL and WL level-shifter logic circuits (Fig. 8).

In order to control and synchronize various units while taking advantage of the eligible time-overlap between consecutive VMM operations as well as steps of one VMM operation, a multi sub-unit controller is proposed as shown in Fig. 7c. In this controller, various duties are delegated to different sub-units. Router handles the data transfer between MM and collector/loader/processor port. Loader is able to perform single/burst read from MM into AUX or buffers with various buffer/MM address strides. Similarly, a collector is able to perform single/burst write from IDUs/AUX into MM with various buffer/MM address strides. The control signals for synchronizing various steps of VMM operation such as selecting 3D-NAND memory layers, applying inputs, performing VMM operation, and calculating outputs are issued by operator. Finally, the main controller configures the rest of the sub-units and synchronize their operations. Such controller design enables reading the data for the next VMM operation, while writing back the output results from the previous one. It also eliminates heavily nested loops (e.g. convolution tasks) in the machine code through implementing the most frequently used loops in hardware.

### 4.3 Network Mapping

The goal of network mapping is to break down inference computation into a sequence of steps (instructions) and to determine optimal locations for storing kernel weights in VMM arrays and temporary results in the main memory. (The mapping process was also crucial for fine-tuning architectural parameters, e.g. understanding minimal requirements for main memory capacity.)

To do that, the neural network is first converted into computational graph in which each node represents one (convolution, fully-connected, max-pooling, etc.) network layer, while each edge represents the amount of data which has to be transferred from one node (layer) to others. The layers are processed sequentially as a sequence of "processing steps" and we assume that all input and output data of the currently processed layer has to be stored in memory. With such scheme, the total amount of main memory which will be occupied after each processing step can be calculated by counting the edges in the computational graph which are cut by a line separating all already processed nodes from yet-to-be processed ones. Fig. 9a shows the memory requirement graph extracted from such assessment performed for the studied networks.

The weight matrices are mapped into the 3D structure of memory blocks using a weight placement scheme including three steps - namely quantization, reshaping, and 3D packing. According to this scheme, first, the weight kernel dimensions, i.e. number of inputs and outputs, are quantized by $K$. In convolution operation, the quantization, reshaping, and packing are performed in such a way that the shift operation in hardware is equivalent to the shift in convolution. Then, in the second step, the quantized weight matrix dimensions are compared to the maximum dimensions of one-step VMM in the hardware, i.e. $2N \times M$. If the kernel dimensions exceed the maximum allowable 2D VMM in hardware in any dimension, the weight matrix is broken in that dimension and reshaped to a 3D matrix in such a way that the third dimension, which is equivalent to the memory layer in a hardware, indicates different weight sub-matrices (either in a row-first or column-first manner).

In a third step, weight kernels are mapped into specific locations in 3D memory array using heuristic algorithm whose goal is to minimize the number of utilized memory cell layers. Specifically, one iteration of the algorithm involves generation of a randomly ordered list of kernels, and then sequential mapping of kernels from the list by greedily searching for the locations within already occupied memory layers, and only allocating new layers if no such location is found. The best solution is then chosen among several iterations of the algorithm. The output results of such algorithm are shown in Fig. 9b for the three studied networks.

## 5. SYSTEM-LEVEL PERFORMANCE

In order to evaluate the system-level performance for any target DNN/RNN network running on the 3D-aCortex, we have developed a software framework that utilizes the post-layout energy/speed/area metrics of all its blocks (buffers, buses, DTCs, TDCs, neurons, and digital circuits) in the 55-nm technology node. (The energy/throughput/area numbers for the SRAM-based instruction memory and the eDRAM-based main memory are obtained using the Cacti memory estimator [45].) This framework extracts the list of processing tasks for a given network, maps the VMM kernels on the 3D array of memory devices, and provides estimates for the energy/throughput of the inference operation along with the area of the processor for the given set of architecture specifications.

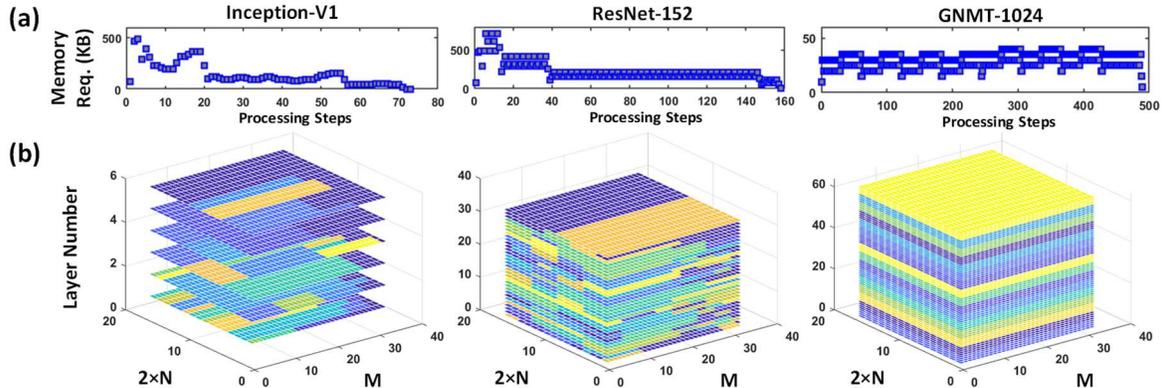

**Fig. 9:** (a) Memory requirement for various processing steps of a single inference task, and (b) mapping of the network weight kernels into the 3D-NAND memory block (different colors represent different kernels, and dark blue represents empty space) for Inception-v1 [4], ResNet-152 [5], and GNMT-1024 [6].

Two DNN networks, Inception-v1 [4] and ResNet [5], with different computational graphs and network sizes, and also Google's natural machine translation (GNMT), a very common RNN network [6], have been selected as the benchmarks for the evaluation of the proposed general-purpose architecture. The evaluation was performed for 3D-aCortex with 4-bit computing (activation) precision, which seems to be sufficient for the studied networks. For example, Refs. 51, 52 reported negligible drop in functional performance compared to the full precision one for exactly the same version of ResNet which was studied in our work, a larger version of Inception, and, similar to GNMT, LSTM-based recurrent networks.

Furthermore, we have performed a preliminary exploration of architectural parameters to optimize the processor's performance. As a result, the value $K = 64$ was chosen to improve the computational block utilization, while the parameters $M$ and $N$ were selected to balance the read and write time/energy. Note that the parasitics of the shared bit lines (O-BUS) bound the horizontal dimension of the processor, and it, in turn, affects the PE's aspect ratio and the number ($2N$) of these elements sharing one line. A detailed study of the benchmark networks has shown that a 1MB MM is sufficient to store all intermediate data, while the flow control program requires at most 4KB IM. Finally, our rigorous analysis indicates that $M = 32$ and $N = 8$ satisfies the aforementioned conditions while being sufficient to perform even the largest, 128M-weight benchmark GNMT.

The architecture specifications, performance measures and their breakdown are summarized in Fig. 7e for the baseline 3D-aCortex. Fig. 8b,c shows a brief report, focusing on the most important differences, for shared-CAP architecture.

## 6. COMPARISON WITH PRIOR WORK

On the circuit level, to the best of our knowledge, a 3D-NAND-based VMM has been studied in only one work [39].

The main assumption of that work was that the word line of every cell in a particular layer is partitioned into separate independent lines along $x$-direction. However, such modification would require major changes in the fabrication flow of the existing 3D-NAND memory technology. This approach also faces a very challenging problem of managing a large number of word lines, which would likely result in a very heavy peripheral overhead. In addition, in this scheme, based on the current-mode VMM, analog input signals are applied to highly resistive and capacitive word lines, leading to higher energy consumption and larger delays.

In contrast, our approach is fully compatible with commercial 3D-NAND flash memory. Encoding of inputs by digital pulses that are applied to bit-select lines results in better energy-efficiency and speed. Moreover, as it was shown in the prior work on the VMMs based on the 2D floating-gate memory [27], the time-domain approach to the VMM task enables more compact and energy-efficient peripheral circuits than the usual current mode implementations. As a result, as the detailed simulations described in Sec. 3 have shown, the energy efficiency of our 3D-VMM is very high - for example, ~9 fF/Op for $M = N$ up to 500, at the 5-bit accuracy. Our results also show that the proposed 3D-VMM achieves a ~100× area efficiency increase in comparison with its 2D-NOR memory-based counterpart [27], while maintaining a comparable energy efficiency and throughput. Moreover, digital nature of the circuit peripheries (level-shifters, neuron, DTC, and TDC) in our proposed time-domain VMM significantly relaxes the limitation for technology node scaling as opposed to the analog nature of the peripheral circuits in the amplifier/current-mirror (voltage/current-mode) based approaches. In the proposed design, the precision is mainly constrained by inherent flash device characteristics such as DIBL and capacitive coupling (and not the peripheral circuitry characteristics such as gain, noise, and their sensitivity to process variation). Considering the extremely small footprint of the flash cells due to 3D integration, the proposed approach can significantly benefit from technology node scaling even with scaling limitation of the flash cells.

| Platform | DaDianNao [8] | TPU [7] # | UNPU[14] # | ISAAC [46] | PUMA [47] | 2D-aCortex [18] | 3D-aCortex & |
|---|---|---|---|---|---|---|---|
| Technology node | 28 nm | 28 nm | 65 nm | 32 nm | 32 nm | 55 nm | 55 nm |
| Approach | digital | digital | digital | ReRAM | ReRAM | 2D-NOR | 3D-NAND |
| Clock (MHz) | 606 | 700 | 200 | 1200 | 1000 | 700 | 1000 |
| Precision (bits) | 16 fixed point | 8 fixed point | 1-16 (4 here) | 16 fixed point | 16 fixed point | 4 fixed point | 4 fixed point |
| Area (mm²) | 88 | 330 | 16 | 85.4 | 90.6 | 292.9 | 18.43 / 41.7 |
| Power (W) | 20.1 | 40 | 297 | 65.8 | 62.5 | 0.039 | 0.151 / 0.126 |
| Throughput (TOps/s) | 5.54 | 92 | 1.38 | 39.9 | 52.31 | 14.97 | 10.66 / 8.2 |
| CE (TOps/s-mm²) | 0.063 | 0.28 | 0.086 | 0.46 (0.62*) | 0.58 (0.78*) | 0.051 | 0.58 / 0.2 |
| SE (MB/mm²) | 0.2 | off-chip | off-chip | 0.74 (0.25*) | 0.76 (0.257*) | 0.273 | 4.34 / 30.7 |
| EE (TOps/J) | 0.286 | 0.43 | 11.6 | 0.35 (5.14*) | 0.84 (12.09*) | 380.25 | 70.43 / 65 |

**Table 2: Performance comparison of 3D-aCortex to the state-of-the-art digital and mixed-signal neuromorphic processor architectures. Except for TPU and UNPU, all performance results are based on simulations. *Estimated, highly optimistic performance for 4-bit computing precision and 55-nm technology node implementation. # The performance numbers do not include overhead of external memory access (weights/intermediate data). & Baseline / 16x CAP sharing architectures. 4-bit weight/activation quantization results in negligible decrease in functional performance (and actually better performance for ResNet) [51,52].**

On the other hand, on the system level quite a few efforts were recently made to exploit the efficiency of MS operators to develop better DNN/RNN processor architectures [46-52]. For example, the ISAAC [46] and PUMA [47] architectures are 2D mesh structures of tiles where each tile contains several small (typically 128×128) ReRAM-based VMM units with their I/O peripheries. In these architectures, one shared memory is implemented in each tile for storing intermediate data and communication between the VMMs, while communications between the tiles are performed through a shared 2D bus structure. Such heavily-granular, multi-core design approach aims at increasing the VMM unit utilization, minimizing the data transfer overhead, and maximizing the system throughput via pipelining and parallel processing. However, the data conversion / communication overhead due to the partial VMM operation, static power consumption and large area overhead of the neurons / DACs / ADCs, and a large control and communication overhead between tiles / VMMs likely limits the performance of such architectures, especially when running relatively complex computational graphs such as those of the Inception [4] and ResNet [5] tasks.

In contrast to this prior system-level work, our 3D-aCortex processor architecture is harmonically matched with the proposed 3D-NAND VMM as the core processing unit. It includes a flexible/programmable granular single-bank 3D analog operator and a reconfigurable folded chain of buffers, which allows contingent implementation of various size VMMs and convolution kernels fully in time/analog domain. Such design results in maximizing the data reuse while minimizing the area overhead of peripheral and control circuitry, as well as the energy overhead of the VMM operation (integration and I/O conversion) and control/data-movement associated with heavily multi-core designs performing partial VMM operations [46,47]. The main advantages of the proposed architecture are:

- A flexible single-bank design, which results in very large sharing factor of costly peripheral circuitry such as buffers,

DTCs, neurons, TDCs, and programming circuitry, while maintaining the capability of performing various size VMM operations. The large sharing factor of the peripheral circuitry and a high density of 3D-NAND memory result in a remarkable storage efficiency.

- Such a design provides a flexible large VMM operator fully in time/analog domain, and consequently allows contingent implementation of VMMs of various size, fully exploiting the energy efficiency and speed of computation in time/analog domain, i.e. avoiding overheads of partial VMM operations.

- The layer-by-layer processing scheme, combined with the single-bank deployment of analog operators, result in a relatively simple control circuitry, with low energy/area overhead, while still supporting even complex computational graphs.

- The data reuse in convolution layers is fully preserved via a configurable folded buffer chain design.

- Due to the time-domain approach, zero static power of the computational blocks improves the energy efficiency.

The detailed simulation results for 3D-aCortex, benchmarked on representative RNN/DNN models, have shown a performance significantly higher than all published prior results, including the fully digital [7, 8, 14] and MS [18, 46, 47] systems - especially for mobile/IoT applications, for which the storage and energy efficiencies are the most important metrics (Table 2). In order to make a fair comparison between 3D-aCortex and other MS approaches, we have performed a highly optimistic rescaling of the published performance metrics to the 55-nm, 4-bit design point. Even with this highly optimistic projection, the baseline and shared-CAP 3D-aCortex provides a ~17× / ~119× improvement of the storage efficiency, and a ~14× / ~ 13× improvement of the energy efficiency over the ISAAC [46], while maintaining a comparable computational efficiency of 0.58 / 0.2 TOps/(s-mm²). In comparison with PUMA [47], these numbers are, respectively ~17× / ~119× and ~6× / ~5.5×. These results also show that in comparison with the 2D-aCortex based on 55-nm NOR flash memory [18], the chip footprint of the 3D a-Cortex is ~16 / ~7 times smaller, while its energy efficiency is lower only a factor of ~5.4 / ~ 5.

Moreover, the proposed 3D-aCortex architecture is based on the 3D-NAND flash technology and digital time-domain peripheral circuitry, allowing for its further scaling beyond 20-nm technology node without performance/precision degradation. This fact promises even more compact and energy-efficient neuromorphic processors based on future, more advanced technology nodes.